\documentclass[aps,prl,article,twocolumn,showpacs,preprintnumbers,amsmath,amssymb,superscriptaddress]{revtex4}

\date{\today}
\usepackage{epsfig}
\usepackage{subfigure}
\usepackage{graphicx}% Include figure files
\usepackage{dcolumn}% Align table columns on decimal point
\usepackage{bm}% bold math
\usepackage[colorlinks,linkcolor=blue,hyperindex,CJKbookmarks]{hyperref}
\usepackage{float}
\usepackage{hyperref}
\usepackage{comment}
\newcommand{\tJ}{$t\small{-}J$}

\newcommand{\tJs}{$t\small{-}J\small{-} 3s$}
\newcommand{\tJzs}{$t\small{-}J_z\small{-} 3s$}

\newcommand{\Akw}{$A({\bf k},\omega)$}

\newcommand{\tpr}{$t^\prime$}

\begin{document}

\title{Origin of Strong Dispersion in Hubbard Insulators}
\author{Y. Wang }
\altaffiliation{Y. Wang and K. Wohlfeld contributed equally to this work. Correspondence should be addressed to:
Y. Wang (yaowang@stanford.edu) or K. Wohlfeld (krzysztof.wohlfeld@fuw.edu.pl).}
\affiliation{Department of Applied Physics, Stanford University, California 94305, USA}
\affiliation{Stanford Institute for Materials and Energy Sciences, SLAC National Accelerator Laboratory and Stanford University, Menlo Park, CA 94025, USA}
\author{K. Wohlfeld}%
\altaffiliation{Y. Wang and K. Wohlfeld contributed equally to this work. Correspondence should be addressed to:
Y. Wang (yaowang@stanford.edu) or K. Wohlfeld (krzysztof.wohlfeld@fuw.edu.pl).}
\affiliation{Stanford Institute for Materials and Energy Sciences, SLAC National Accelerator Laboratory and Stanford University, Menlo Park, CA 94025, USA}
\affiliation{Institute of Theoretical Physics, Faculty of Physics, University of Warsaw, Pasteura 5, PL-02093 Warsaw, Poland}
\author{B. Moritz}
\affiliation{Stanford Institute for Materials and Energy Sciences, SLAC National Accelerator Laboratory and Stanford University, Menlo Park, CA 94025, USA}
\affiliation{Department of Physics and Astrophysics, University of North Dakota, Grand Forks, North Dakota 58202, USA}
\author{C.J. Jia}
\affiliation{Stanford Institute for Materials and Energy Sciences, SLAC National Accelerator Laboratory and Stanford University, Menlo Park, CA 94025, USA}
\author{M. van Veenendaal}
\affiliation{Advanced Photon Source, Argonne National Laboratory, Argonne, IL 60439, USA}
\affiliation{Department of Physics, Northern Illinois University, De Kalb, IL 60115, USA}
\author{K. Wu}
\affiliation{Stanford Institute for Materials and Energy Sciences, SLAC National Accelerator Laboratory and Stanford University, Menlo Park, CA 94025, USA}
\author{C.-C. Chen}
\affiliation{Advanced Photon Source, Argonne National Laboratory, Argonne, IL 60439, USA}
\author{T. P. Devereaux}
\affiliation{Stanford Institute for Materials and Energy Sciences, SLAC National Accelerator Laboratory and Stanford University, Menlo Park, CA 94025, USA}

\date{\today}
\begin{abstract}
Using cluster perturbation theory, we explain the origin of the strongly dispersive feature found at high binding energy in the spectral function of the Hubbard model.
By comparing the Hubbard and \tJs\ model spectra, we show that this dispersion does not originate from either coupling to spin fluctuations ($\propto\! J$) or the free hopping ($\propto\! t$). Instead, it should be attributed to a long-range, correlated hopping $\propto\! t^2/U$, which allows an effectively free motion of the hole within the same antiferromagnetic sublattice. This origin explains both the formation of the high energy anomaly in the single-particle spectrum and the sensitivity of the high binding energy dispersion to the next-nearest-neighbor hopping $t^\prime$.
\end{abstract}
\pacs{71.10.Fd, 74.72.Cj, 79.60.-i}

\maketitle
\section{Introduction}
High-temperature superconductivity in the cuprate oxides has attracted significant attention over the past thirty years. However, the precise origin of this phenomenon is not well understood due to the complex physics even in a minimal model used to describe the correlated nature of the electrons~\cite{anderson1987resonating,zhang1988effective, PhysRevLett.61.1415}:
the 2D Hubbard model. It is often assumed that a first step in understanding the physics of this model is to study its spectral properties~\cite{Damascelli2003, RevModPhys.70.1039, PhysRevB.44.10256, Dagotto1992, PhysRevB.47.1160, Preuss:1995SDW, Groeber2000, Moritz:2009fb, Senechal:2000fg,Senechal:2002fr,senechal2012book,pairault1998strong, Kohno:2012PRL, PhysRevB.90.035111} in the simple undoped limit.

\begin{figure}[!t]
\begin{center}
\includegraphics[width=\columnwidth]{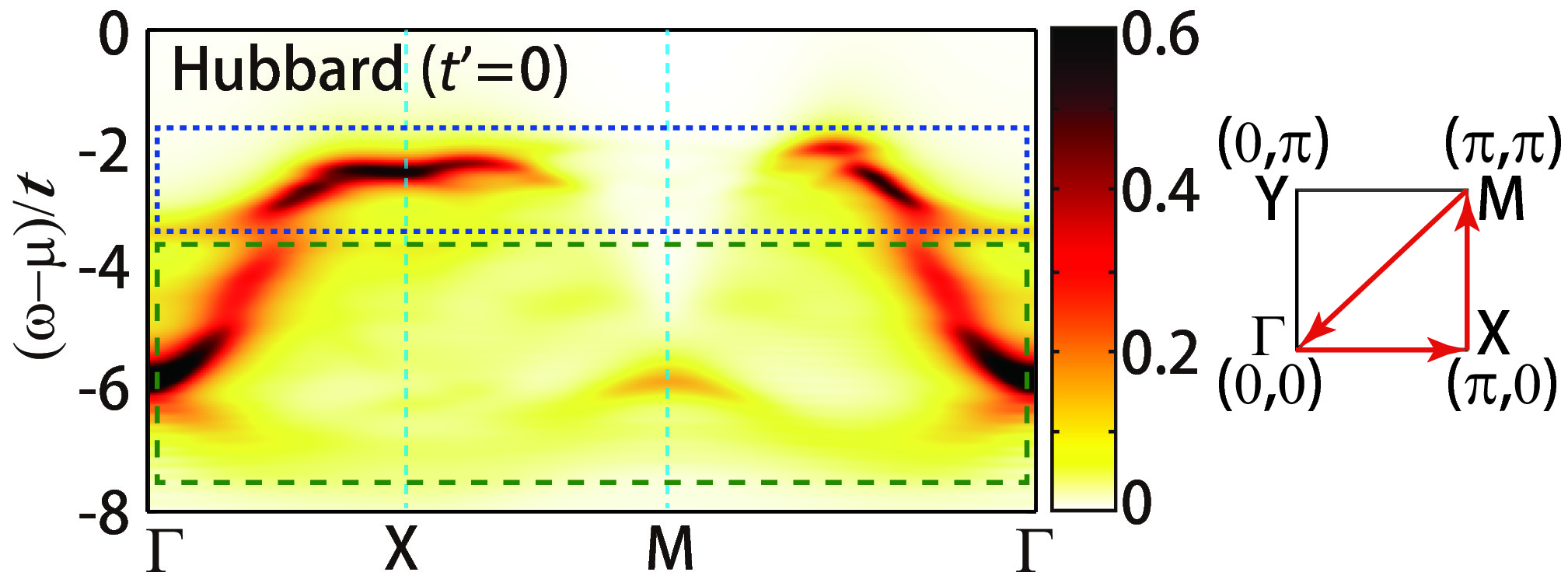}
\caption{\label{fig:1}(color online)
Left panel: spectral function \Akw\ of the undoped Hubbard model with $U=8t$ and $t^\prime=0$ calculated using CPT on a 4$\times$4 cluster for the high symmetry directions of the 2D Brillouin zone.
The blue dotted (green dashed) box guides the eye for the LBE (HBE) features.
Right panel: the high symmetry directions of the 2D Brillouin zone.
}
\end{center}
\end{figure}

The spectral function of the undoped 2D Hubbard model when the free electron bandwidth $W$
is comparable to the Hubbard interaction $U$ consists of two prominent features in the band structure, cf.~Fig.~\ref{fig:1}. At low binding energies (LBE), a sharply defined quasiparticle-like excitation disperses downward from $(\pi/2,\pi/2)$ reaching an energy of the order of spin exchange $J=4t^2/U$ near the $\Gamma$-point (0,0). This quasiparticle, often labeled the spin polaron (SP)~\cite{Martinez:1991SpinPolaron}, represents a hole heavily dressed by spin excitations from the antiferromagnetic (AF) ground state with its bandwidth no longer governed by the free electron hopping $t$ but by the spin exchange $J$~\cite{Schmitt:1988SCBAThry, Kane:1989SCBAThry, Martinez:1991SpinPolaron, Efstratios:2007SCBA, Zemljic:2008, Preuss:1995SDW, macridin2007high}.

%%%%%%%%%%%%%%%%%%%%%%%%%%%%%%%%%%%%%%%%%%%%%%%%%%%%%%%%%

%%               FIGURE 1
%%%%%%%%%%%%%%%%%%%%%%%%%%%%%%%%%%%%%%%%%%%%%%%%%%%%%%%%%

At higher binding energies (HBE), another feature becomes prominent approaching the $\Gamma$-point. The delineation of the LBE spin polaron from the HBE feature has been widely associated with the ``high energy anomaly'' or ``waterfall'' seen in a number of cuprate photoemission results~\cite{MottwaterfallExp, waterfallExp, kordyuk2005bare,meevasana2007hierarchy,valla2007high,chang2007low,xie2007high,pan2006universal, Moritz:2009fb, Byczuk:2007en, Efstratios:2007SCBA, macridin2007high, Zemljic:2008, PhysRevB.84.205107, Maekawa:2001ui, tohyama1996approximate, inosov2007momentum, greco2007t, greco2008self, foussats2008role, markiewicz2007paramagnon}. While the spin polaron is well understood, the characterization remains poor for the feature at higher energies. One may naively expect that a broad, non-dispersive Hubbard band should appear at the $\Gamma$-point. However, as shown in Fig.~\ref{fig:1}, it is clear that there is a sharp dispersion. Previous studies have attributed this feature to scenarios such as spin-charge separation~\cite{Kohno:2012PRL, PhysRevB.90.035111, Kohno:2010hf, Kohno:2014JPS, Maekawa:2001ui, tohyama1996approximate, maekawa1997electronic, maekawa2000spin, weng2001spin} or a weak-coupling spin-density wave~\cite{Bulut:1994SDW,Preuss:1995SDW, Groeber2000}. However, these interpretations remain controversial.

The aim of this paper is to understand the nature of the HBE feature and what separates it from the SP. Using cluster perturbation theory we examine both the Hubbard, \tJ\, and \tJs\ models to provide an in-depth examination of what controls the quasiparticle dispersion and intensity at low and high energies. We demonstrate that the HBE dispersion is mainly set by a correlated effective hopping $\sim\!t^2/U$
rather than the free hopping $t$.
While the SP dispersion is set by spin exchange $J$ due to the strong coupling of the mobile hole to spin excitations,
the HBE dispersion is determined by hopping within the same AF sublattice through long-range (so-called 3-site) correlated hopping.
Therefore, one naturally expects a distinct transition between them, providing a physical picture for the high energy anomaly.
Moreover, as the uncorrelated next-nearest-neighbor (nnn) hopping $t^\prime$ also allows for hopping within the same AF sublattice,
we obtain a natural explanation for the sensitivity of the Hubbard model spectrum at high energies to changes in $t^\prime$.

%%%%%%%%%%%%%%%%%%%%%%%%%%%%%%%%%%%%%%%%%%%%%%%%%%%%%%%%%

%%               FIGURE 2
%%%%%%%%%%%%%%%%%%%%%%%%%%%%%%%%%%%%%%%%%%%%%%%%%%%%%%%%%
\begin{figure}[!t]
\begin{center}
\includegraphics[width=0.76\columnwidth]{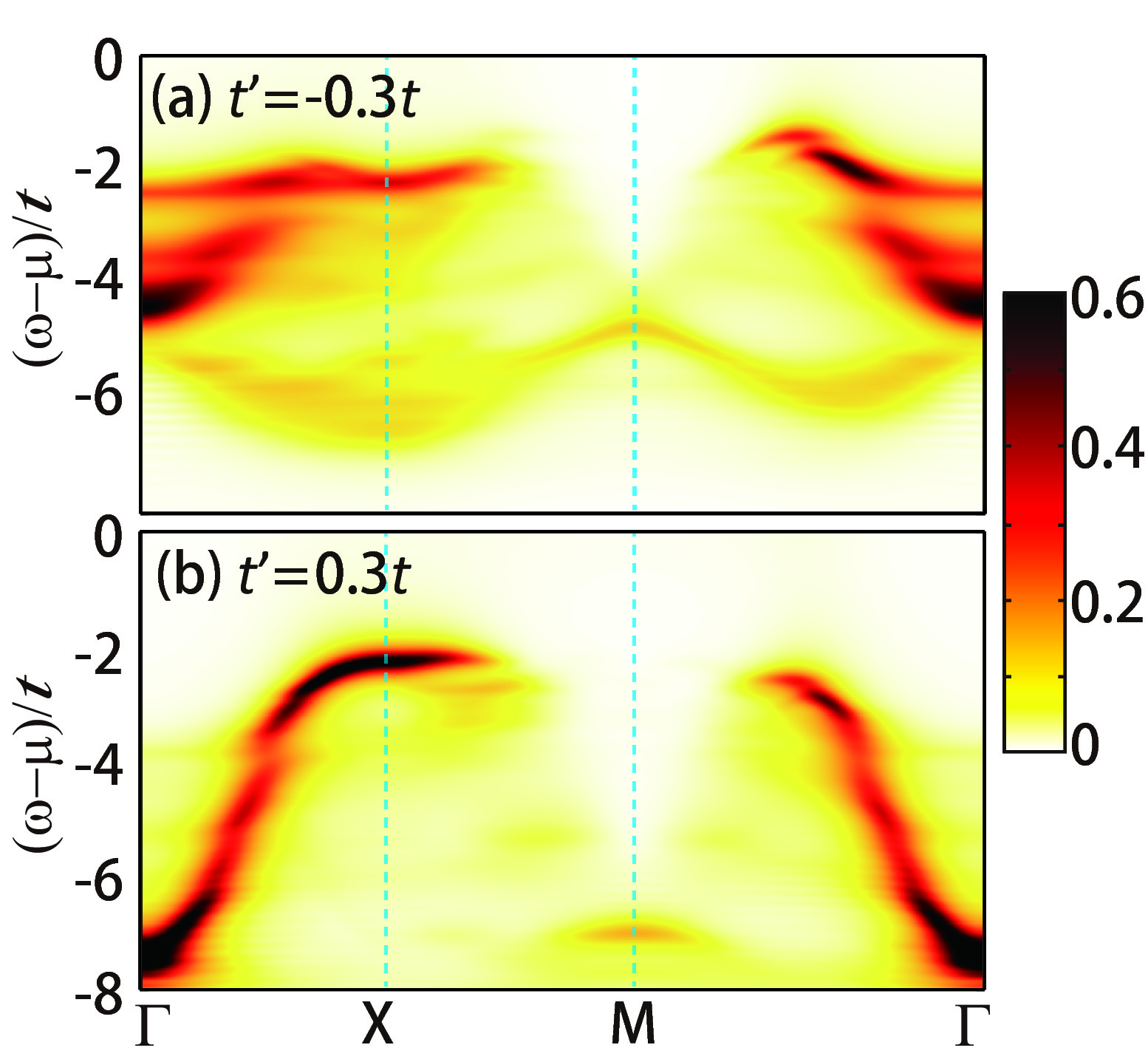}
\caption{\label{fig:2}(color online) Spectral function \Akw\ for the undoped Hubbard model with $U=8t$ and (a) $t^\prime=-0.3t$ and (b) $t^\prime=0.3t$ calculated using CPT on a 4$\times$4 cluster.
}
\end{center}
\end{figure}

\section{Models and Methods}
The 2D Hubbard model Hamiltonian $\mathcal{H}_{H}$ is
\begin{eqnarray} \label{eq:hubbard}
\mathcal{H}_{H}&=&-\sum_{{\bf i},{\bf j},\sigma}  \left(t_{\bf ij} c^\dagger_{{\bf j} \sigma} c_{{\bf i} \sigma} +h.c.\right) \nonumber\\
&&+ U\sum_{{\bf i}}\left(\!n_{{\bf i}\uparrow}-\frac12\right)\left(n_{{\bf i}\downarrow}-\frac12\right),
\end{eqnarray}
where $c^\dagger_{{\bf i}\sigma}$ ($c_{{\bf i}\sigma}$) and $n_{{\bf i}\sigma}$ denotes the creation (annihilation) and density operator at site $\textbf{i}$ of spin $\sigma$; $t_{\bf ij}$ is the hopping, restricted here to nearest-neighbors (nn) $t_{\bf \langle ij\rangle}\!=\!t$ and nnn $t_{\bf \langle\!\langle ij\rangle\!\rangle}\!=\!t^{\prime}$.
The Hubbard spectral function has been calculated using various numerical methods, \textit{e.g.} exact diagonalization (ED)~\cite{PhysRevB.44.10256, Dagotto1992, RevModPhys.70.1039}, quantum Monte Carlo (QMC)~\cite{PhysRevB.47.1160, Preuss:1995SDW, Groeber2000, Moritz:2009fb} or cluster perturbation theory (CPT) based on ED~\cite{Senechal:2000fg,Senechal:2002fr,senechal2012book,pairault1998strong, Kohno:2012PRL, PhysRevB.90.035111}.
Although CPT is an approximate method, we believe that for the purpose of this paper it is most suitable, as it can produce continuous momentum resolution at zero-temperature.

It is well known that $t^\prime$ qualitatively changes the quasiparticle dispersion and mimics the differences between hole- and electron-doping in cuprates found using Hamiltonians which incorporate higher energy degrees of freedom~\cite{MottwaterfallExp, waterfallExp, kordyuk2005bare,meevasana2007hierarchy,valla2007high,chang2007low,xie2007high,pan2006universal, Moritz:2009fb, Byczuk:2007en, Efstratios:2007SCBA, macridin2007high, Zemljic:2008, PhysRevB.84.205107, Maekawa:2001ui, tohyama1996approximate}, cf.~Fig.~\ref{fig:2}. A positive (negative) \tpr\ is found to increase (decrease) the high-energy anomaly which shows that we can indeed identify two distinct features
(LBE and HBE) in the spectrum.
Moreover, this comparison also shows that the HBE dispersion is far more affected by \tpr\ than is the SP and that changing \tpr\ does not introduce new qualitative features in the spectral function.
This suggests that the dispersion should somehow be related to
an effective nnn hopping which must be present even when $t^\prime\!=\!0$.

%%%%%%%%%%%%%%%%%%%%%%%%%%%%%%%%%%%%%%%%%%%%%%%%%%%%%%%%%

%%               FIGURE 3
%%%%%%%%%%%%%%%%%%%%%%%%%%%%%%%%%%%%%%%%%%%%%%%%%%%%%%%%%
\begin{figure}[t!]
\begin{center}
\includegraphics[width=\columnwidth]{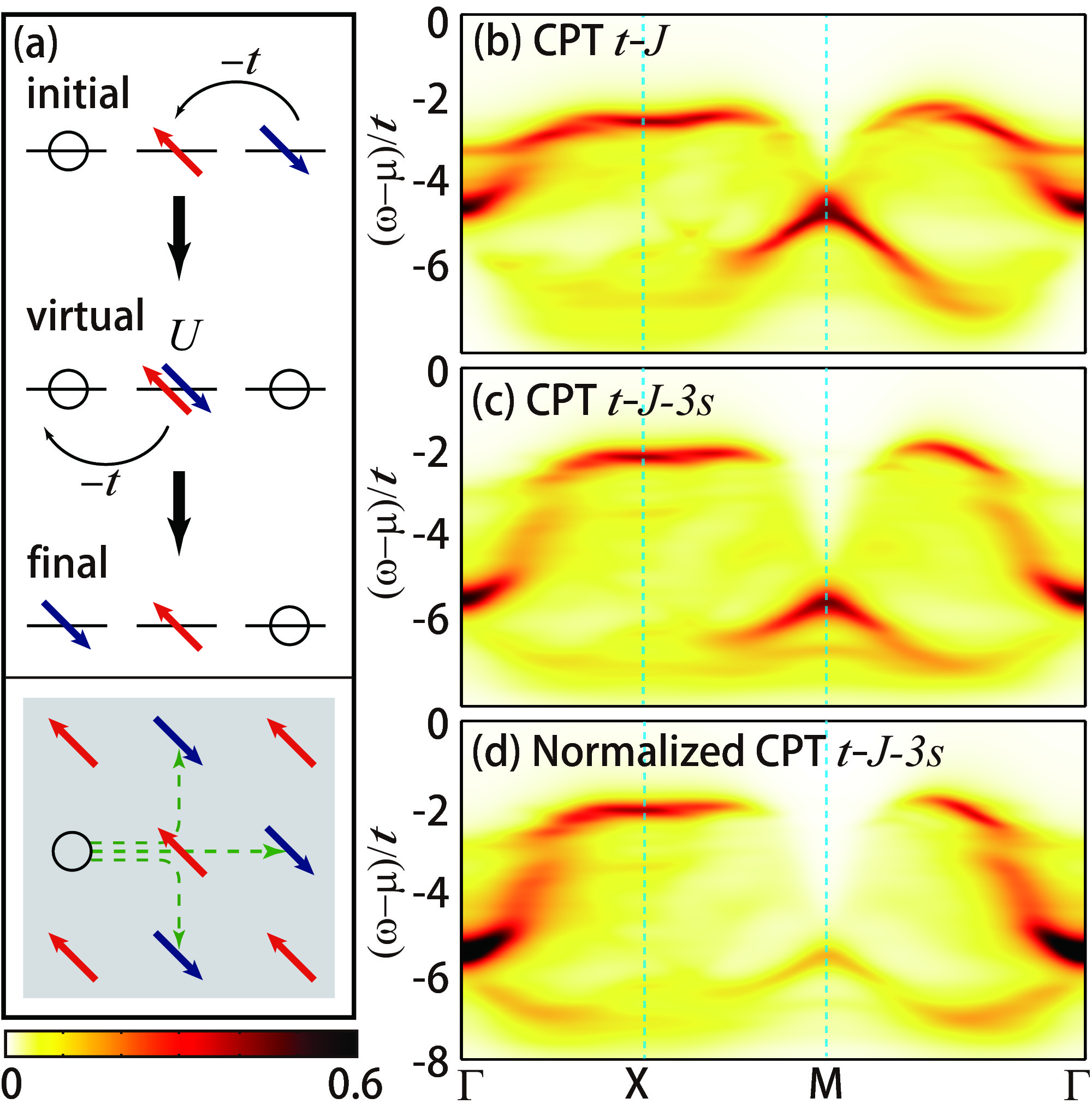}
\caption{\label{fig:3}(color online) (a) Upper panel: schematic cartoon showing the 3-site correlated hopping elements $\propto t^2/U$ in an exchange process via virtual double occupancy $\propto U$; lower panel: various 3-site paths in the AF background. Arrows (circles) denote electron spins (holes).
(b-c) Spectral function \Akw\ for the undoped (b) \tJ\ and (c) \tJs\ model with $J=0.5t$ calculated using CPT on a 4$\times$4 cluster. (d) The same as (c) but with the spectral weight normalized in the Hubbard--like way, see text.
}
\end{center}
\end{figure}

While this
%% correlated
effective long-range
hopping is not explicitly present in the Hubbard model when $t^\prime\!=\!0$, it does appear in a $t/U$ expansion to lowest order~\cite{chao1977tJmodel, chao1978canonical, belinicher1994consistent, belinicher1994range}
as the so-called 3-site term, see Fig.~\ref{fig:3}(a) and Refs.~\onlinecite{stephan1992single, tJ3s, Bala:1995ttprJSCBA, Spalek1988, Szczepanski1990, Eskes1994, Eskes1994b, belinicher1994consistent, belinicher1994range, belinicher1996generalized, belinicher1996single, psaltakis1992optical, Eskes1996, e2g3s, PhysRevB.78.214423:3site, Kuzmin2014}.
This defines the \tJs\ model with the Hamiltonian given by $\mathcal{H}_{t\!-\!J\!-\!3s}= \mathcal{H}_{t \!-\!J}+\mathcal{H}_{3s}$:
\begin{align}\label{tJH}
\mathcal{H}_{t\!-\!J}&=-t\!\sum_{\langle {\bf i},{\bf j} \rangle,\sigma}\! \left( \tilde{c}^\dagger_{{\bf j} \sigma} \tilde{c}_{{\bf i}\sigma}\! +\!h.c.\right)\!+\!J\sum_{\langle {\bf i},{\bf j}\rangle}\left( \textbf{S}_{\bf i} \! \cdot \! \textbf{S}_{\bf j}-\frac{n_in_j}{4}\right)\!, \nonumber\\
\mathcal{H}_{3s}&= - \frac{J_{3s}}{4}\!\!\sum_{\langle {\bf i},{\bf j}\rangle,\langle {\bf i},{\bf j}^\prime\rangle\atop {\bf j}\neq {\bf j}^\prime,\sigma}\! \! \left(\tilde{c}^\dagger_{{\bf j}^\prime\sigma}\tilde{n}_{{\bf i}\bar{\sigma}}\tilde{c}_{{\bf j}\sigma}\!+ \! \tilde{c}^\dagger_{{\bf j}^\prime\sigma} \tilde{c}^\dagger_{{\bf i}\bar{\sigma}} \tilde{c}_{{\bf i}\sigma} \tilde{c}_{{\bf j}\bar{\sigma}}\right)\!,
\end{align}
where $\textbf{S}_{\bf i}\cdot\textbf{S}_{\bf j}\! =\! S_{\bf i}^zS^z_{\bf j}\!+\!\frac12\left(S_{\bf i}^+S_{\bf j}^-+S_{\bf i}^-S_{\bf j}^+\right)$, with $S_{\bf i}^z\!=\!(n_{{\bf i} \uparrow}\!-\!n_{{\bf i}\downarrow})/2$ and $S_{\bf i}^+\!=\!(S_{\bf i}^-)^\dagger\!=\!\tilde{c}^\dagger_{{\bf i}\uparrow} \tilde{c}_{{\bf i}\downarrow}$.
The constrained fermionic operators acting in the Hilbert space without double occupancies are defined as $\tilde{c}^{\dag}_{{\bf i}\sigma} = {c}_{{\bf i}\sigma}^{\dag}(1-n_{{\bf i} \bar{\sigma}})$. Note that only when $\!J_{3s}\!=J$ does this \tJs\ model follow from a perturbative expansion of the Hubbard model, Eq.~(\ref{eq:hubbard}).

\section{Spectral Features}
To verify
%%this postulate,
that the 3-site terms are responsible for the onset of the HBE dispersion,
we calculate the spectral function of both the \tJ\ and \tJs\ model using CPT, see Figs.~\ref{fig:3}(b) and (c) respectively~\cite{Note1}.
At low energies the \tJ\ and \tJs\ model spectra are quite similar, see Refs.~\onlinecite{Bala:1995ttprJSCBA, Ebrahimnejad2015}, and qualitatively reproduce the SP dispersion in the Hubbard spectrum. The qualitative agreement extends to the HBE only in the \tJs\ model, showing explicitly that the 3-site terms~\cite{tJ3s, Bala:1995ttprJSCBA, Spalek1988, Szczepanski1990, Eskes1994, Eskes1994b, Eskes1996, e2g3s, PhysRevB.78.214423:3site, Kuzmin2014} indeed play a crucial role in the development of HBE dispersion. We note that the spectral weight of the \tJs\ and Hubbard models remains quite different which can be attributed to different sum rules. Artificially introducing equivalent spectral weight sum rules between the two models [see Fig.~\ref{fig:3}(d)] produces an even more qualitative and quantitative agreement between the two spectra (see Appendix \ref{appA} for a detailed discussion).

%%%%%%%%%%%%%%%%%%%%%%%%%%%%%%%%%%%%%%%%%%%%%%%%%%%%%%%%%

%%               FIGURE 4
%%%%%%%%%%%%%%%%%%%%%%%%%%%%%%%%%%%%%%%%%%%%%%%%%%%%%%%%%
\begin{figure}[h!]
\begin{center}
\includegraphics[width=\columnwidth]{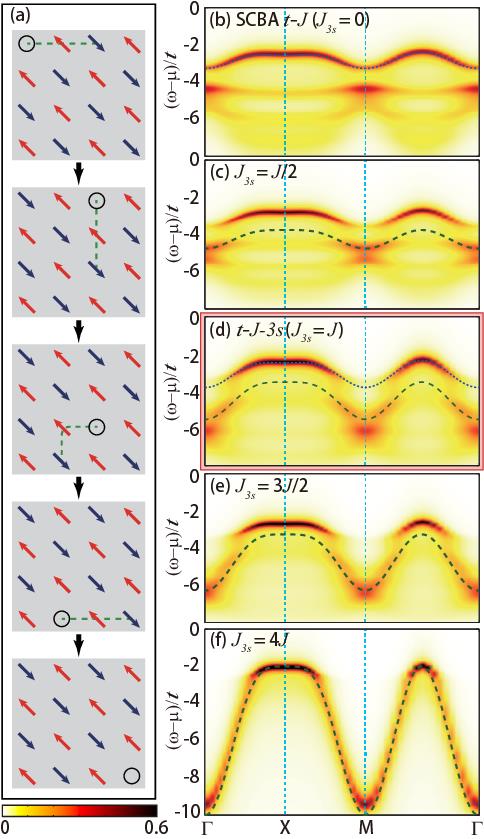}
\caption{\label{fig:4}(color online)
(a) Idealized propagation via IASH (dashed lines) triggered by \textit{e.g.} the 3-site terms: the hole motion does not disturb the AF background.
(b-f) Spectral functions \Akw\ for various 3-site term values in \tJs\ models using SCBA on a 40$\times$40 cluster ($J=0.5t$ and broadening $\delta=0.15t$), with $J_{3s}=0$, $J/2$, $J$, $3J/2$ and $4J$, respectively. [The canonical parameter $J_{3s}=J$ is highlighted by the red box, panel (d)].
Dotted lines indicate fitted SP dispersion, while dashed lines indicate IASH dispersion Eq.~(\ref{IASHeq}).
}
\end{center}
\end{figure}

With a correlation established between the HBE dispersion and the 3-site terms in a $t/U$ expansion of the Hubbard model, a natural question is whether or not the dispersion is real and, if so, how does one understand the underlying mechanism which dominates this dispersion. We postulate that the 3-site terms open an additional channel through which a hole can ``freely'' propagate within the same AF sublattice. This motion does not disturb the underlying AF order, as shown intuitively in Fig.~\ref{fig:4}(a), and we call this mechanism an intra--AF-sublattice effective hopping (IASH). The dispersion relation is given by
\begin{equation}\label{IASHeq}
\varepsilon_{{\bf k}}^{\rm IASH} \!=\! -\frac{U}{2}\!-\! \frac{J_{3s}}{2}\left[\cos 2k_x \!+\!\cos 2k_y \!+\!4\cos k_x \cos k_y \right],
\end{equation}
which can be deduced from the $\mathcal{H}_{3s}$ term assuming a perfect AF background.

To test this postulate and provide an answer to these questions, we calculate \Akw\ for the \tJs\ model using a more analytical technique -- the self-consistent Born approximation (SCBA)~\cite{Martinez:1991SpinPolaron, Bala:1995ttprJSCBA, Brink1998, Efstratios:2007SCBA}.
This approximate technique is not limited to small cluster sizes~\cite{Martinez:1991SpinPolaron} and can be viewed as complementary to CPT.
To emphasize the role of the 3-site terms, we calculate the \tJs\ model using SCBA for various values of $J_{3s}$ as shown in Figs.~\ref{fig:4}(b-f), where the physical value of $J_{3s}$ obtained from the perturbative expansion of Hubbard model is $J_{3s}=J$ \cite{stephan1992single, tJ3s, Bala:1995ttprJSCBA, Spalek1988, Szczepanski1990, Eskes1994, Eskes1994b, belinicher1994consistent, belinicher1994range, belinicher1996generalized, belinicher1996single, psaltakis1992optical} [see Fig.~\ref{fig:4}(d)].
First, with increasing $J_{3s}$ a continuous spectral weight transfer occurs from the top of the SP to higher energies, proceeding from the $\Gamma$ and M points~\cite{Bala:1995ttprJSCBA, Efstratios:2007SCBA, macridin2007high, Zemljic:2008}.
Next, comparing Fig.~\ref{fig:4}(d) with Fig.~\ref{fig:3}(c), we conclude that the HBE dispersion in both CPT and SCBA shows a very specific shape around the $\Gamma$ point with the ``proper'' value of the 3-site terms. This suggests that such a dispersion is ``real'' and not an artifact of the CPT method which artificially may have enhanced the dispersion due to the approximation. Neglecting the 3-site terms [Fig.~\ref{fig:4}(a)] suppresses the signatures of dispersion at high energy~\cite{Note2} with the only prominent dispersion associated with the SP (see Appendix \ref{appB} and Refs.~\onlinecite{Martinez:1991SpinPolaron, Bala:1995ttprJSCBA, Efstratios:2007SCBA} for a detailed discussion). As a single dispersive branch is visible for an artificially large value of $J_{3s}$ [see Fig.~\ref{fig:4}(f)], there is likely an onset of the dispersion due to the 3-site terms.

The HBE dispersion always is well-approximated by Eq.~(\ref{IASHeq}) [cf.~Fig.~\ref{fig:4}(b-f)]:
for smaller $J_{3s}$ values this is visible around the $\Gamma$ and M points and the region of agreement grows across the Brillouin zone for larger $J_{3s}$. Note that this dispersion relation is deduced from $\mathcal{H}_{3s}$ assuming a perfect AF background and neglecting the coupling to quantum spin fluctuations. This suggests that in the corresponding parts of the Brillouin zone the hole indeed can be mobile, unencumbered by spin fluctuations, as illustrated by the cartoon in Fig.~\ref{fig:4}(a). This is qualitatively different from the situation in the \tJzs\ model, as discussed in the Appendix \ref{appC} and Refs.~\onlinecite{e2g3s, PhysRevB.78.214423:3site}.

\section{Conclusions and Discussion}
In summary, we have presented a detailed understanding of the spectral function of the undoped Hubbard model, identifying distinct origins for the low and high binding energy dispersions. Comparing the spectral functions of the Hubbard, \tJ\, and \tJs\ models, we have argued that the dispersion at high binding energy originates from an effective intra--antiferromagnetic-sublattice hopping, primarily due to the so-called 3-site terms. We found that this dispersion relation is not renormalized by coupling to spin fluctuations which suggests that holes at these energy scales can propagate more-or-less freely on the same sublattice. This distinction between the physical mechanisms which give rise to the dispersion at both low and high energies, and thus the character of the associate quasiparticles, unveils a rather profound origin for the high energy anomaly.
The result also has important consequences for the Hubbard model with uncorrelated nnn hopping $t^\prime$, explaining its crucial impact on the dispersion primarily at high binding energy.

Given the clear dichotomy between the features of the spectral function at low and high energy and the correlated, but effectively free, hopping associated with the 3-site terms, we might expect distinct effects on emergent behavior, \textit{e.g.}~antiferromagnetism, charge order, superconductivity, or some combination as found in the cuprates. In particular, since $t^\prime$ or nnn hopping has been linked to $T_c$~\cite{pavarini2001band,chen2009unusual}, this may further indicate a causal connection between the HBE feature and material dependence of $T_c$. Moreover, the fact that a doped hole can effectively decouple from the antiferromagnetic background and move without disrupting the spin order in part of the Brillouin zone is consistent with the recently observed persistence of magnetic excitations upon doping\cite{le2011intense, dean2013persistence, lee2014asymmetry, ishii2014high, jia2014persistent, wang2014real}.
 Since none of the prominent dispersing features in the half-filled Hubbard model are due to the free hopping $t$, a natural follow-up question would be when and how one might expect the uncorrelated hopping $t$ to become important upon doping? Since there should be a crossover between the Mott insulator near half-filling and a weakly-correlated metal at substantial doping, the answer to this question could shed some light on our understanding of
superconductivity which appears near optimal doping in the cuprates.

\section*{ACKNOWLEDGEMENTS}
We thank J. van den Brink, Y. He, A. M. Ole{\'s}, Z.-X. Shen, and J. Spa\l{}ek for insightful discussions. This work was supported at SLAC and Stanford University by the US Department of Energy, Office of Basic Energy Sciences, Division of Materials Sciences and Engineering, under Contract No.~DE-AC02-76SF00515 and by the Computational Materials and Chemical Sciences Network (CMCSN) under Contract No.~DE-SC0007091 for the collaboration. Y.W. was supported by the Stanford Graduate Fellows in Science and Engineering. C.C.C. is supported by the Aneesur Rahman Postdoctoral Fellowship at Argonne National Laboratory (ANL), operated by the US Department of Energy (DOE) Contract No. DE-AC02-06CH11357. M.v.V. is supported by the DOE Office of Basic Energy Sciences (BES) Award No. DE-FG02-03ER46097 and the NIU Institute for Nanoscience, Engineering and Technology. K.Wohlfeld acknowledges support from the Polish National Science Center (NCN) under Project No. 2012/04/A/ST3/00331.
A portion of the computational work was performed using the resources of the National Energy Research Scientific Computing Center supported by the US Department of Energy, Office of Science, under Contract No.~DE-AC02-05CH11231.

\setcounter{figure}{0}
\renewcommand\thefigure{A\arabic{figure}}
\appendix
\section{Details of the CPT Calculations}\label{appA}
{\it Evaluation of the spectral function}
The single particle spectral function of any quantum mechanical Hamiltonian $\mathcal{H}$ is defined as:
\begin{equation}\label{eq:akw}
A({\bf k},\omega)=-\frac1{\pi}\textrm{Im} \sum_\sigma\langle G|c^\dagger_{{\bf k}\sigma}\frac1{\omega+\mathcal{H}-E_G+i\delta}c_{{\bf k} \sigma}|G\rangle,
\end{equation}
where $c^\dagger_{{\bf k} \sigma}$ ($c_{{\bf k} \sigma}$) denotes the electron creation (annihilation) operator with momentum ${\bf k}$ and spin $\sigma$,
$|G\rangle$ is the ground state with energy $E_G$, and $\delta$ is a Lorentzian broadening.

On a finite cluster, the Hamiltonian can be split into $\mathcal{H}=\mathcal{H}_c +\mathcal{H}_{\rm int}$, where $\mathcal{H}_c$ contains the (open-boundary) intra-cluster terms
while $\mathcal{H}_{\rm int}$ contains the inter-cluster hopping terms. The CPT method evaluates $\mathcal{H}_{c}$ exactly using exact diagonalization, and treats $\mathcal{H}_{\rm int}$ perturbatively.
The corrected Green function can be expressed as
\begin{equation}\label{1storder}
\mathcal{G}=\frac{G_c}{1-\mathcal{H}_{\rm int}G_c},
\end{equation}
where $G_c$ is the cluster Green function (evaluated from $\mathcal{H}_c$). In the long-wavelength limit
\begin{equation}
\mathcal{G}_{CPT}(\textbf{k},\omega)=\frac1{N}\sum_{a,b}\mathcal{G}_{a,b}(\textbf{k},\omega)e^{i\textbf{k}\cdot(\textbf{r}_a-\textbf{r}_b)},
\end{equation}
where $a,b$ are intra-cluster site indices.
Under the CPT method, the spectral function is evaluated by
\begin{equation}
A(k,\omega)=-\frac1\pi \mathcal{G}_{CPT}(\textbf{k},\omega).
\end{equation}

{\it Justification of using CPT}
Although CPT is not strictly applicable to \tJ-like models because the two-particle vertex (spin exchange on the boundary) is not well defined,
the $t$ vertex should be dominant in an intra-cluster $t/U$ expansion of the Hubbard model.
Furthermore, for these quantum cluster approaches the key to the quality of the results lies in the cluster sizes,
instead of higher-order corrections to the CPT formula.
We also have studied the $t\small{-}J\small{-}3s$ model on various cluster sizes such as 20 and 24 sites (not shown) and find that the dispersion saturates with increasing cluster size. In addition, one can assume that the agreement between the Hubbard and $t\small{-}J\small{-}3s$ model spectra (Figs.~\ref{fig:1} and \ref{fig:3}) cannot be a coincidence (compared over different cluster sizes), which provides an {\it a posteriori} justification that the CPT method with only first order $t$-vertex can be acceptable for the $t\small{-}J$ models.

%%%%%%%%%%%%%%%%%%%%%%%%%%%%%%%%%%%%%%%%%%%%%%%%%%%%%%%%%

%%               FIGURE S1
%%%%%%%%%%%%%%%%%%%%%%%%%%%%%%%%%%%%%%%%%%%%%%%%%%%%%%%%%
\begin{figure}[!t]
\begin{center}
\includegraphics[width=6.5cm]{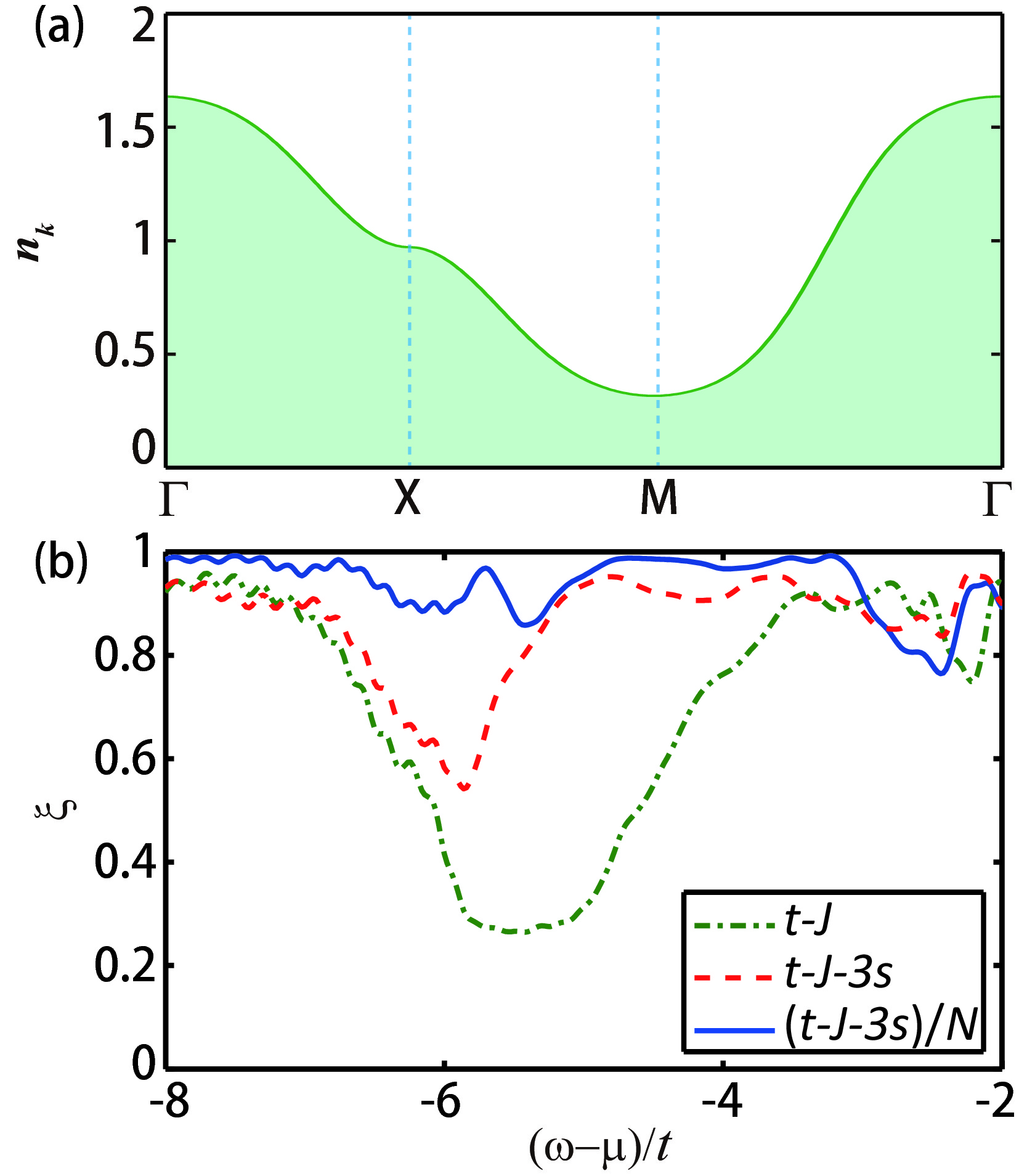}
\caption{\label{fig:s1}(color online) (a) The electron density $n_{\bf k}$ used for normalization in Fig.~\ref{fig:3}. (b) The momentum-integrated correlation function $\xi(\omega)$ [Eq.~(\ref{eq:ksiw})] calculated for the CPT \Akw\ results for: the \tJ\ and Hubbard models (dot-dashed lines), \tJs\ and Hubbard models (dashed lines) and \tJs\ models normalized in the Hubbard-like way [or (\tJs)$/N$] and Hubbard models (solid lines).
}
\end{center}
\end{figure}

{\it ``Normalization'' of the \tJs\ spectra}
Figure~\ref{fig:s1}(a) shows the momentum distribution of integrated spectral weight (electron density) $n_{\bf k}$ in the lower Hubbard band (below E$_\mathrm{F}$). This is different in the \tJs\ model, where due to the projection of the upper Hubbard band (forbidden double-occupancy), the spectrum has a uniform $n_{\bf k}$ distribution. Thus, to partially correct for this difference and enhance comparisons, the Fig.~\ref{fig:3}(d) is obtained by dividing the spectrum found in Fig.~\ref{fig:3}(c) by the $n_{\bf k}$ distribution for the Hubbard model [Fig.~\ref{fig:s1}(a)] at each momentum.

{\it Quantitative comparison of CPT results}
In what follows we show in a quantitative way that: (i) the CPT results for the \tJs\ model are closer to those of the Hubbard model
than the CPT results for the ``bare'' \tJ\ model, and (ii) the CPT results for the \tJs\ model, normalized in a Hubbard--like way, qualitatively reproduce the Hubbard model spectral function extremely well (\textit{i.e.}~better than unrenormalized \tJs\ results).

We define the (momentum-integrated) correlation function of two spectral functions $A^{(\alpha)}(\textbf{k},\omega)$ and $A^{(\beta)}(\textbf{k},\omega)$:
\begin{equation}\label{eq:ksiw}
\xi^{(\alpha\beta)}(\omega)=\frac{\int A^{(\alpha)}(\textbf{k},\omega)A^{(\beta)}(\textbf{k},\omega)d^2\textbf{k}}{\sqrt{\int|A^{(\alpha)}(\textbf{k},\omega)|^2d^2\textbf{k}\int|A^{(\beta)}(\textbf{k},\omega)|^2d^2\textbf{k}}}.
\end{equation}

Fig.~\ref{fig:s1}(b) shows the correlation function between the Hubbard spectrum, with \tJ\ , \tJs\ , and normalized \tJs\ spectrum calculated via CPT. One concludes that: (i) indeed the ``normalized'' \tJs\ model reproduces extremely well the spectrum of the Hubbard model also on a {\it semi-quantitative} level, (ii) the ``unnormalized'' \tJs\ model provides a marginally ``worse'' comparison than the ``normalized'' \tJs\ model, and (iii) the ``bare'' \tJ\ model certainly is unable to capture the features, especially at high binding energy ($\sim\ \omega-\mu<-4t$), of the Hubbard model.

\setcounter{figure}{0}
\setcounter{equation}{0}
\renewcommand\thefigure{B\arabic{figure}}
\renewcommand\theequation{B\arabic{equation}}
\section{Details of the SCBA Calculations}\label{appB}
To calculate the spectral function \Akw\ for the undoped \tJs\ model, we first assume that the ground state has (i) a broken spin rotational symmetry and (ii) long range AF order. These two assumptions allow us to map the \tJs\ model onto the so-called spin-polaron Hamiltonian, cf.~Eq.~(2.15) in Ref.~\onlinecite{Bala:1995ttprJSCBA}. Next the spectral function for the spin polaron Hamiltonian is calculated by using the SCBA: we calculate the self-energy by summing all noncrossing diagrams to the infinite order and self-consistently evaluate the single particle Green's function on 40$\times$40 lattice sites.

\setcounter{figure}{0}
\renewcommand\thefigure{C\arabic{figure}}
%%%%%%%%%%%%%%%%%%%%%%%%%%%%%%%%%%%%%%%%%%%%%%%%%%%%%%%%%

%%               FIGURE S2
%%%%%%%%%%%%%%%%%%%%%%%%%%%%%%%%%%%%%%%%%%%%%%%%%%%%%%%%%
\begin{figure}[!h]
\begin{center}
\includegraphics[width=0.65\columnwidth]{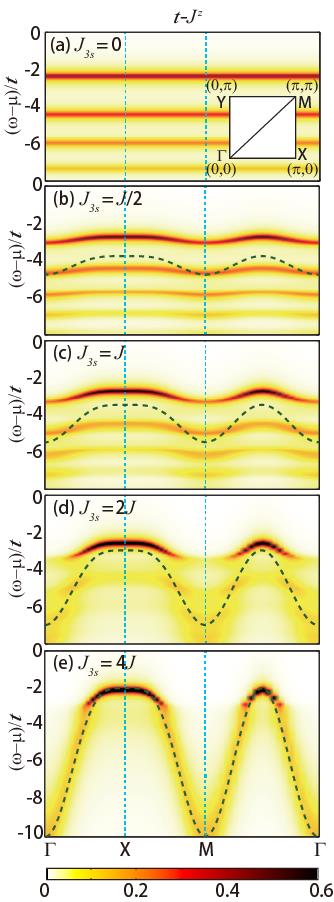}
\caption{\label{fig:s2}(color online)
Comparison of \Akw\ for \tJzs\ model calculated using SCBA for various strength of the 3-site term $J_{3s}$:
(a) $J_{3s}=0$, (b) $J_{3s}=J/2$, (c) $J_{3s}=J$, (d) $J_{3s}=2J$ and (e) $J_{3s}=4J$, see text for further details. The dashed line indicates the IASH dispersion following from the inclusion of the 3-site terms in the model.
}
\end{center}
\end{figure}

We also present here the functional form of the SP dispersion relations following the considerations of Ref.~\onlinecite{Martinez:1991SpinPolaron}:
\begin{eqnarray}
\varepsilon_{{\bf k}}^{\rm SP} &=&x_1 + x_2 (\cos k_x + \cos k_y)^2\nonumber\\
&& + x_3 \left[\cos(2k_x)+\cos(2k_y)\right],
\end{eqnarray}
where $x_1$, $x_2$, and $x_3$ are parameters which depend on the ratio of $J/t$ in the \tJ\--like model and which follow from fitting the above dispersion relation to the dispersion of the upper edge of the continuum in the SCBA calculated spectra.
In the case of the \tJ\ model we obtain $x_1 =-1.5t$, $x_2 =0.18t$, and $x_3 = 0.035t$ (note a slightly different constant shift $x_1$ with respect to the one adopted in Ref.~\onlinecite{Martinez:1991SpinPolaron} due to a different system size in the SCBA calculations). In the case of SCBA calculations of the \tJs\ model we obtain $x_1 =-1.6t$, $x_2 =0.33t$, and $x_3 = 0.035t$.
Note that the difference between the fitting parameters for \tJ\ and \tJs\ models suggests that the SP dispersion relation is influenced by the presence of the 3-site terms. This suggests that the hole forming the SP moves not only as a result of coupling to the spin fluctuations but also via the 3-site terms.

\section{Influence of Quantum Fluctuation on the IASH}\label{appC}
We calculate the spectral function \Akw\ using SCBA for the \tJzs\ model
with varying strength of the 3-site term where

\begin{eqnarray}\label{tJzH}
\mathcal{H}_{t\!-\!J_z-\!3s}&=&-t\!\sum_{\langle {\bf i},{\bf j} \rangle,\sigma}\! \left( \tilde{c}^\dagger_{{\bf j} \sigma} \tilde{c}_{{\bf i}\sigma}\! +\!h.c.\right)\!+J\sum_{\langle i,j\rangle}\left[S^z_{\bf i} S^z_{\bf j}-\frac{n_in_j}{4}\right] \nonumber \\
&& -\frac{J_{3s}}{4}\!\!\sum_{\langle {\bf i},{\bf j}\rangle,\langle {\bf i},{\bf j}^\prime\rangle\atop {\bf j}\neq {\bf j}^\prime,\sigma}\! \! \tilde{c}^\dagger_{{\bf j}^\prime\sigma}\tilde{n}_{{\bf i}\bar{\sigma}}\tilde{c}_{{\bf j}\sigma}\!.
\end{eqnarray}

Compared with Fig.~\ref{fig:4}, the result presented in Fig.~\ref{fig:s2} shows that when quantum spin fluctuations are switched off the IASH shows up only for unrealistically large strength of the 3-site hopping term.
As shown in Fig.~\ref{fig:s2}, the spectrum of the \tJzs\ model shows almost no sign of an {\it unrenormalized} free dispersion which should result from the inclusion of the 3-site terms. Instead, each of the polaronic--like states, visible in the ladder spectrum, acquires a renormalized dispersion due to the onset of the 3-site terms. In this case the IASH and the polaronic--like propagation are not independent of one another, as already explained in detail some time ago in Refs.~\onlinecite{e2g3s, PhysRevB.78.214423:3site}. It is then only in the case of an unrealistically large value of the 3-site terms that the unrenormalized dispersion appears in the spectral function of the \tJzs\ model.

This discrepancy between the \tJzs\ and \tJs\ models is in contrast with the interesting result of Ref.~\onlinecite{Ebrahimnejad:2014em}.
In that case the reported irrelevance of quantum spin fluctuations on the hole motion in a charge transfer insulator (as governed by the three-band model) concerns solely the effective SP dispersion and is due to the rather similar dispersion relation of the quantum-spin-fluctuation--mediated hopping and the IASH.
\bibliography{paper}
\clearpage
\end{document}